\documentclass[
superscriptaddress,
twocolumn,
amsmath,amssymb,
longbibliography,
aps,
prx,
showpacs
]{revtex4-1}
\usepackage{graphicx}% Include figure files
\usepackage{dcolumn}% Align table columns on decimal point
\usepackage{bm}% bold math
\usepackage{color}
\usepackage[colorlinks,linkcolor=red,citecolor=blue]{hyperref}
\usepackage{xcolor}
\usepackage{braket}

\newcommand{\br}[1]{\pmb{\mathrm{#1}}}

\newcommand{\mg}[1]{\lvert {#1} \rvert}

\begin{document}
	
	%\preprint{APS/123-QED}
	
	\title{Broadband strong optical dichroism in topological Dirac semimetals with \\ Fermi velocity anisotropy}% Force line breaks with \\
	
	\author{J. Lim}
	\affiliation{%
		Science, Math and Technology, Singapore University of Technology and Design,  Singapore
	}%
	
	\author{K. J. A. Ooi}
	\affiliation{
		School of Energy and Chemical Engineering, Xiamen University Malaysia, Selangor Darul Ehsan 43900, Malaysia}
	\affiliation{
		College of Chemistry and Chemical Engineering, Xiamen University, Xiamen 361005, China
	}
	
	\author{C. Zhang}
	\affiliation{%
		School of Physics, University of Wollongong, Northfields Avenue, New South Wales 2522, Australia
	}%
	
	\author{L. K. Ang}
	\email{ricky\_ang@sutd.edu.sg}
	\affiliation{%
		Science, Math and Technology, Singapore University of Technology and Design, Singapore
	}%
	
	\author{Yee Sin Ang}
	\email{yeesin\_ang@sutd.edu.sg}
	\affiliation{%
		Science, Math and Technology, Singapore University of Technology and Design,  Singapore
	}%

	%===============================================
	% abstract (max 150 words)
	%===============================================
	\begin{abstract}
		Prototypical three-dimensional (3D) topological Dirac semimetals (DSMs), such as Cd$_3$As$_2$ and Na$_3$Bi, contain electrons that obey a linear momentum-energy dispersion with different Fermi velocities along the three orthogonal momentum dimensions. Despite being extensively studied in recent years, the inherent \emph{Fermi velocity anisotropy} has often been neglected in the theoretical and numerical studies of 3D DSMs. Although this omission does not qualitatively alter the physics of light-driven massless quasiparticles in 3D DSMs, it does \emph{quantitatively} change the optical coefficients which can lead to nontrivial implications in terms of nanophotonics and plasmonics applications. Here we study the linear optical response of 3D DSMs for general Fermi velocity values along each direction. Although the signature conductivity-frequency scaling, $\sigma(\omega) \propto \omega$, of 3D Dirac fermion is well-protected from Fermi velocity anisotropy, the linear optical response exhibits strong linear dichroism as captured by the \emph{universal} extinction ratio scaling law, $\Lambda_{ij} = (v_i/v_j)^2$ (where $i\neq j$ denotes the three spatial coordinates $x,y,z$, and $v_i$ is the $i$-direction Fermi velocity), which is independent of frequency, temperature, doping, and carrier scattering lifetime. For Cd$_3$As$_2$ and Na$_3$Bi$_3$, an exceptionally strong extinction ratio larger than 15 and covering broad terahertz window is revealed. Our findings shed new light on the role of Fermi velocity anisotropy in the optical response of Dirac semimetals and open up novel polarization-sensitive functionalities, such as photodetection and light modulation.  
		
	\end{abstract}

	
	\maketitle
	
	\section{Introduction}
	In recent decades, significant effort has been dedicated to the study of topological insulators (TIs) and low-dimension semimetals like graphene. In the low-energy limit, the electrons occupying these exotic materials obey relativistic, Dirac-like hamiltonians. Their behaviour as pseudo-massless particles confers these materials unique optical properties such as high EM field confinement~\cite{Koppens,Graphene_Plasmonics,Chen2012,Fei2012} and strong optical nonlinearity~\cite{wright,PhysRevB.95.125408,PhysRevB.82.201402,Mikhailov_2007,PhysRevLett.105.097401,ang2010nonlinear, shareef2012room, ang2012enhanced, chen2012photomixing, ang2015nonlinear, matt, huang2019strong}, making them desirable for uses like table-top generation of high-brightness coherent radiation spanning from the X-ray to the terahertz regimes through mechanisms such as free-electron-graphene plasmon scattering~\cite{Wong2016a,Rosolen2018}, high-harmonic generation~\cite{Yoshikawa736,Cox2017,THzHHG_hotcarriers, lee2015negative}, and transition radiation~\cite{OE_Zhang17}, and also as saturable absorbers for infrared ultrafast lasers~\cite{Sun2010,Popa2010,BaoQ2009,Zhang2012}.

	Recently, a new class of quantum materials which behave like bulk analogues of graphene have also attracted significant attention -- 3D Dirac semimetals (DSMs). Unlike their TI counterparts which possess only conducting surface states, 3D DSMs are also conducting in the bulk. 
	The dispersion of 3D DSMs are formed from a superposition of two Weyl cones of opposite chirality~\cite{PhysRevB.85.195320,BurkovAA_NatureMaterials2017}. Each Weyl cone is linearly dispersing in all three momentum directions and is doubly-degenerate at a single band-touching point (Weyl node). Hence, the energy bands are doubly-degenerate for all momenta except at the Dirac point, where a four-fold degeneracy arises from the overlap of two Weyl nodes.  Topologically unprotected Dirac points may occur at a quantum critical point in the phase transition between a TI and a normal insulator~\cite{PhysRevB.76.205304,NJP_murakami,PhysRevB.78.165313,Xu560,Sato2011,Yang2014,NJP_WeiZhang} or between weak and strong TIs~\cite{PhysRevX.4.011002}. Some Dirac points are guaranteed by virtue of crystal symmetries~\cite{PhysRevLett.108.140405,PhysRevB.85.195320,PhysRevLett.112.036403,Yang2014} and unlike graphene, are robust against spin-orbit interaction-induced gapping. These stable Dirac points have been predicted in materials like $\mathrm{BiO_{2}}$~\cite{PhysRevLett.108.140405} and $\mathrm{A_{3}Bi}$ (A = K, Rb)~\cite{PhysRevB.85.195320}, and experimentally detected in $\mathrm{Cd_{3}As_{2}}$~\cite{Liu2014a,Neupane2014,Borisenko2014} and $\mathrm{Na_{3}Bi}$~\cite{Science_Na3Bi_discovery,na3bi}. 
	
	As the electrons in 3D DSMs also possess linear dispersions, they have been expected to exhibit qualitatively-similar field response to graphene~\cite{APLphotonics_YeeSin,THzHHG_hotcarriers}. Hence, for applications where the properties of graphene are desired in bulk materials, 3D DSMs present themselves as natural candidates~\cite{Liu2014a,APLphotonics_YeeSin,Science_Na3Bi_discovery}.  For instance, $\mathrm{Cd_{3}As_{2}}$ \cite{transport,cd3as2, dirac_transport} has been shown to perform well as saturable absorbers in the mid-IR regime~\cite{Zhu2017} owing to its strong broadband light-matter interaction~\cite{Wang2017,Meng2018}, much like graphene.  However, they can be realized as optical thin films~\cite{Meng2018,Pan2015}, enabling application of conventional semiconductor methods for parameter control~\cite{Meng2018}.  Recent exciting experiments have also demonstrated generation of terahertz radiation up to the $3^{\mathrm{rd}}$~\cite{Cd3As2THzHHG} and $7^{\mathrm{th}}$~\cite{Kovalev2019} harmonics, and a theoretical study predicted generation of harmonics beyond the $31^{\mathrm{st}}$ order~\cite{JLim_3D_DSM_THz_HHG_2020} with conversion efficiencies far exceeding high-harmonic generation (HHG) in graphene by virtue of a finite interaction volume. For these reasons, 3D DSMs are attractive candidates for novel nanophotonic and nanoplasmonic applications or as viable alternatives to graphene. With increased exploration into the potential applications of 3D DSMs, a theoretical understanding of how Dirac physics determine the linear and nonlinear optical properties is crucial. 
	The slope of the linearly dispersing Dirac conic band structure around the nodal band touching point is commonly referred as the \emph{Fermi velocity}, which serves as an important velocity-like parameter that characterizes the electronic band structure of Dirac materials.
	While a few works~\cite{PhysRevB.93.235417,APLphotonics_YeeSin,PhysRevB.93.085426} have provided a theoretical treatment for the linear optical response of a 3D Dirac dispersion, none have studied effects of the pronounced anisotropy between the Fermi velocities along different directions -- a property inherently present in 3D DSMs which have been discovered so far, on the linear optical response. As such, it is crucial to include these effects and examine if these anisotropies can be exploited for novel applications. 
	
	In this work, we derive the linear interband and intraband conductivities using the Kubo formula. We use a Hamilonian which describes arbitrary Fermi velocity values in each direction, taking us beyond previous works which were restricted by the isotropic dispersion. In obtaining our expressions for the dynamic conductivities in each direction, we find that both the intraband and interband conductivities of 3D DSMs scale as $\sigma_{ii} \propto v_{i}/(v_{j}v_{k})$, where $i,j,k\in\{x,y,z\}$, $i$ is the direction parallel to the incident light polarization, and $j$ and $k$ are the directions perpendicular to the incident light polarization. Additionally, we show that the anisotropy between the response of the $i$ and $j$ directions exhibits the analytical universal scaling relation for the optical conductivity, $\sigma_{i}/\sigma_{j}\propto (v_{i}/v_{j})^{2}$. 
	\begin{figure*}[t]
		\centering
		\includegraphics[width = 160mm]{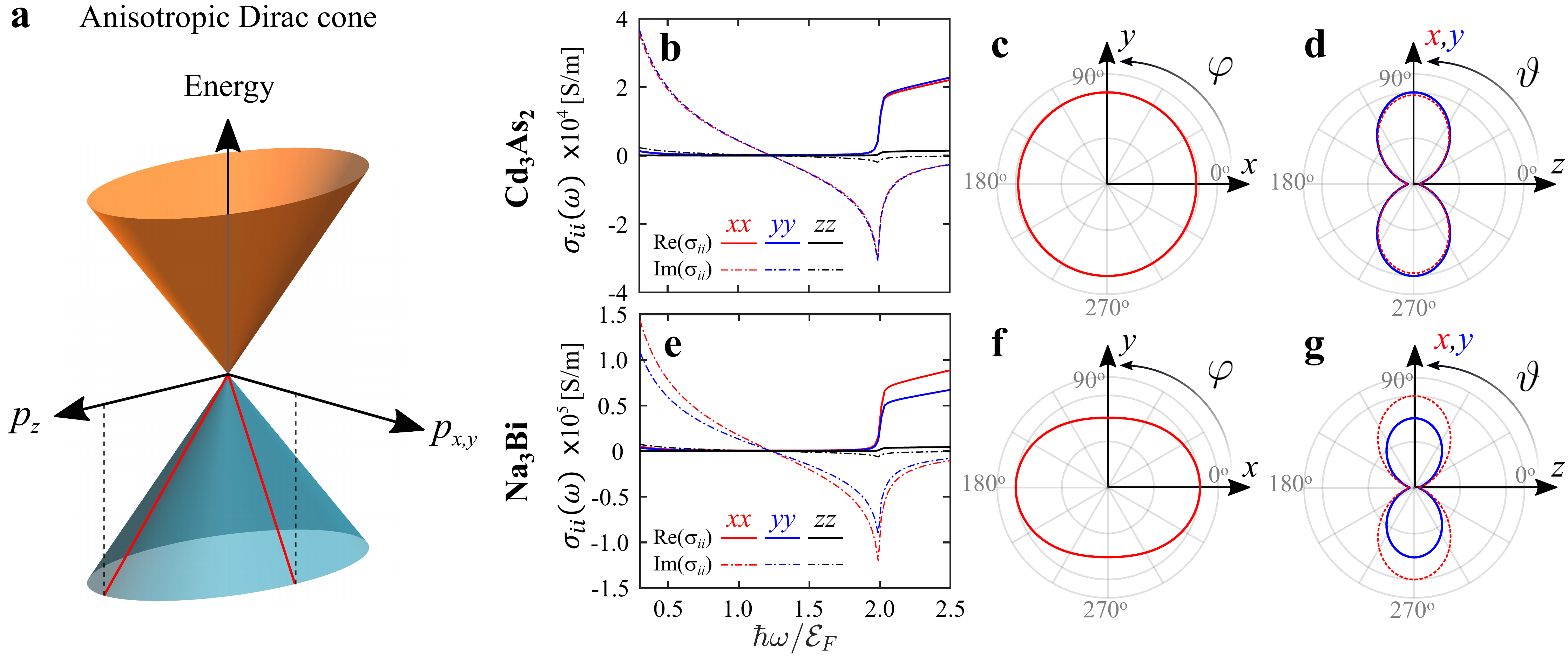}
		\caption{Anisotropic optical response of 3D Dirac semiemtals. (a) Schematic depiction of a 3D Dirac cone banstructure with anisotropic Fermi velocities in each momentum direction. Longitudinal conductivities of anisotropic 3D DSMs $\mathrm{Cd_{3}As_{2}}$ (b) and $\mathrm{Na_{3}Bi}$ (e) in each direction as a function of angular frequency $\omega$. Due to the extreme anisotropy of the Fermi velocities between the in-plane ($p_{x},p_{y}$) and out-of-plane ($p_{z}$) directions in both 3D DSMs,  $\sigma_{zz} \ll \sigma_{xx},\sigma_{yy}$.  The polar plots in (c), (d), (f), and (g) show the directional dependence of the optical conductivity $\sigma$ as a function of angles. For all cases considered here, we considered 3D DSMs doped to $\mathcal{E}_{\mathrm{F}} = 150$ meV at temperature $T = 4$ K, a scattering time of 450 fs, and a combined spin and valley degeneracy $g = 4$. We consider experimentally determined values of the Fermi velocities, as listed in Table~\ref{tbl_anisotropy_ratio} below. }
		\label{Fig_Cd3As2_Na3Bi_sigma_vs_dir}
	\end{figure*}
	
	For realistic 3D DSMs such as Cd$_3$As$_2$ and Na$_3$Bi, this anisotropy ratio can exceed an order of magnitude, emphasizing the importance of including anisotropy in optical calculations. More intriguingly, the \emph{optical extinction ratio} reaches 15.8 and 19.2, respectively, for Cd$_3$As$_2$ and Na$_3$Bi, which is substantially higher than vast majority of anisotropic optical materials previously reported (see Table~\ref{ext_compare} below). Importantly, such optical anisotropy covers a broad terahertz (THz) windows of sub-THz to at least 50 THz and is insensitive to temperature and defect scattering effects.
	Our findings thus reveal the nontrivial role of anisotropic Fermi velocities on the optical response of 3D DSMs and the previously unknown potential of 3D DSM as an exceptionally strong and ultra broadband anisotropic optical material. 
	These results shall form the harbinger for the designs of novel polarization-sensitive nanophotonic, chip-integrable plasmonic and optoelectronic platforms for wide-array of applications such as optical switching, photodetection, energy conversion and light modulation. 
	
	\section{Theory}
	
	Close to the Dirac point, an electron within a 3D Dirac semimetal (DSMs) with momentum $\br{p} = (p_{x},p_{y},p_{z})$ obeys the following low-energy effective Hamiltonian:
	\begin{equation}
	\hat{\mathcal{H}}_{\br{p}} = \sum_{i}v_{i}p_{i}\sigma_{i}
	\end{equation}
	where $i\in\{x,y,z\}$ are the Cartesian components, and $\sigma_{i}$ and $v_{i}$ are the Pauli matrix and the Fermi velocity aligned along direction $i$ respectively. The Hamiltonian admits eigenenergies $\mathcal{E}_{\br{p}} = \pm\sqrt{v_{x}^{2}p_{x}^{2}+v_{y}^{2}p_{y}^{2}+v_{z}^{2}p_{z}^{2}}$, where the positive (negative) branch represents the conduction (valence) band energies. The wavefunction of an electron in band $s\in\{\rm{c},\rm{v}\}$ in the plane wave basis normalized over bulk volume $V$ is:
	\begin{equation}	
	\Psi_{\br{p},s} = \frac{e^{i\br{k}\cdot\br{r}}}{\sqrt{V}}\Phi_{\br{p},s}
	\end{equation}
	where $\br{r}$ is the position vector and $\br{k} = \br{p}/\hbar$ is the crystal momentum vector. The spinor component for band $s$, $\Phi_{\br{p},s}$, reads:
		\begin{subequations}
			\begin{align}
			\Phi_{\br{p},\rm{c}} &= 
			\begin{bmatrix}
			\cos(\theta/2)e^{-i\phi/2}\\
			\sin(\theta/2)e^{+i\phi/2}
			\end{bmatrix}\\
			\Phi_{\br{p},\rm{v}} &= 
			\begin{bmatrix}
			\sin(\theta/2)e^{-i\phi/2}\\
			-\cos(\theta/2)e^{+i\phi/2}
			\end{bmatrix}
			\end{align}
		\end{subequations}
	The angles $\theta = \arccos(q_{z}/q_{r})$ and $\phi = \arctan(q_{y}/q_{x})$ are defined in scaled momentum space coordinates $\br{q} = (v_{x}p_{x},v_{y}p_{y},v_{z}p_{z})$. Note that the radial component $q_{r} = \mg{\mathcal{E}_{\br{p}}}$. The ``c'' (``v'') subscripts denote the conduction (valence) band wavefunctions.  We compute the $i,j$ element of the linear conductivity tensor using the Kubo-Greenwood formula:
	\begin{equation}
	\begin{split}
	\sigma_{ij}(\omega) =& \frac{ge^{2}\hbar}{iV}\sum_{\br{p},\br{p}',s,s'}\Bigg{[}\frac{f_{\mathrm{D}}(\mathcal{E}_{\br{p},s}) - f_{\mathrm{D}}(\mathcal{E}_{\br{p}',s'})}{\mathcal{E}_{\br{p},s} - \mathcal{E}_{\br{p}',s'}}\times\\
	&\qquad\qquad\qquad \frac{\braket{\br{p},s|\hat{v}_{i}|\br{p}',s'}\braket{\br{p}',s'|\hat{v}_{j}|\br{p},s}}{\mathcal{E}_{\br{p},s} - \mathcal{E}_{\br{p}',s'} + \hbar(\omega + i0^{+})}\Bigg{]}
	\end{split}
	\end{equation}
	where $g$ is the combined spin and valley degeneracy, $f_{\mathrm{D}}(\mathcal{E}_{\br{p},s})$ is the Fermi-Dirac distribution, $\hat{v}_{i} = \partial \hat{\mathcal{H}}_{\br{p}}/\partial p_{i} = v_{i}\sigma_{i}$ is the group velocity operator along direction $i$, $\omega$ is the angular frequency of the driving field, and $0^{+}$ is a broadening factor about the pole. The primed (unprimed) variables denote the quantities of the final (initial) state. We concentrate only on the longitudinal ($i = j$) conductivities as the transverse conductivities ($i\neq j$) vanish, leaving only the diagonal tensor terms.  The group velocity expectation values are computed as:
	\begin{equation}
	\braket{\br{p},s|\hat{v}_{i}|\br{p}',s'} = v_{i}\Phi_{\br{p},s}^{\dagger}\sigma_{i}\Phi_{\br{p}',s'}\delta_{\br{k},\br{k}'}= v_{i}\Phi_{\br{p},s}^{\dagger}\sigma_{i}\Phi_{\br{p},s'}
	\end{equation}
	where the $\dagger$ superscript denotes the hermitian conjugate. For the intraband conductivity, we have the general form
	\begin{equation}
	\sigma_{ii}^{\mathrm{intra}} = \frac{ge^{2}v_{i}}{6\pi^{2}\hbar^{3}v_{j}v_{k}}\frac{\tau}{i\omega\tau-1}\int^{+\infty}_{-\infty}\frac{\partial f_{\mathrm{D}}(\mathcal{E})}{\partial\mathcal{E}}\mathcal{E}^{2}d\mathcal{E}
	\label{eqn_intraband_sigma}
	\end{equation}
	where the inelastic scattering time is $\tau$.  The interband conductivity reads:
	\begin{equation}
	\begin{split}
	\sigma_{ii}^{\mathrm{inter}} =& \frac{ige^{2}v_{i}\omega}{3\pi^{2}\hbar^{2}v_{j}v_{k}}\int^{+\infty}_{0}\frac{G(\mathcal{E})\mathcal{E}}{\hbar^{2}(\omega + i0^{+})^{2} - 4\mathcal{E}^{2}}d\mathcal{E},
	\end{split}
	\end{equation}
	where we define the difference between the Fermi-Dirac distributions between both bands as $G(\mathcal{E}) = f_{\mathrm{D}}(-\mathcal{E}) - f_{\mathrm{D}}(\mathcal{E}) = \sinh(\beta\mathcal{E})/[\cosh(\beta\mathcal{E}) + \cosh(\beta\mu)]$ where $\beta = 1/k_{\mathrm{B}}T$, $k_{\mathrm{B}}$ is the Boltzmann constant, and $T$ is the temperature. We can further cast the integral into a form more suitable for numerical calculations:
	\begin{equation}
	\begin{split}
	\sigma_{ii}^{\mathrm{inter}} =& \frac{ge^{2}v_{i}\omega}{3\pi^{2}\hbar^{2}v_{j}v_{k}}\Bigg{\{}i\int^{\mathcal{E}_{\mathrm{c}}}_{0}\frac{G(\mathcal{E}) - G(\hbar\omega/2)}{\hbar^{2}\omega^{2} - 4\mathcal{E}^{2}}\mathcal{E}d\mathcal{E}  \\
	&  \qquad\qquad\qquad+\frac{\pi}{8}G(\hbar\omega/2)\Bigg{\}},
	\end{split}
	\label{eqn_interband_sigma}
	\end{equation}
	where $\mathcal{E}_{\mathrm{c}}$ is the cutoff energy beyond which the band dispersion is no longer linear. We take this values to be $\mathcal{E}_{\mathrm{c}} = 3\mathcal{E}_{F}$~\cite{PhysRevB.93.235417} throughout our work, where $\mathcal{E}_{F}$ is the Fermi level.  To include interband losses, we make the change $\omega \rightarrow \omega+i\tau^{-1}$.  The total conductivity in the $ii$-direction is simply $\sigma^{\mathrm{tot}}_{ii} = \sigma^{\mathrm{inter}}_{ii} + \sigma_{ii}^{\mathrm{intra}}$. In the isotropic limit, our expressions reduce to the results of previous studies. Note that while the $\tau$ in $\sigma_{ii}^{\mathrm{intra}}$ and $\sigma_{ii}^{\mathrm{inter}}$ are generally different, we choose to use the same value for both as there are no qualitative changes to the physics we discuss here. However, one could in principle explicitly derive scattering terms arising as a result of various mechanisms, such as electron-phonon scattering, and long- and short-range impurity scattering.

	\section{Results}
	
	Figure~\ref{Fig_Cd3As2_Na3Bi_sigma_vs_dir} shows the longitudinal conductivity of two recently discovered 3D DSMs  $\mathrm{Cd_{3}As_{2}}$ and $\mathrm{Na_{3}Bi}$ in each Cartesian direction $i\in\{x,y,z\}$. These 3D DSMs possess a strong Fermi velocity anisotropy between the $p_{x}$-$p_{y}$ and $p_{z}$ directions: $v_{z} < v_{x},v_{y}$. Figs.~\ref{Fig_Cd3As2_Na3Bi_sigma_vs_dir}(b) and (e) indicate that the characteristic dependence of the optical conductivity of 3D Dirac fermions on the frequency of the incident light, as seen in previous works~\cite{PhysRevB.93.235417}, remain invariant with respect to direction.  However, the differences in the magnitude of $\mathrm{Re}(\sigma_{ii})$ and $\mathrm{Im}(\sigma_{ii})$ as a function of direction is drastic. The nontrivial difference in the strength of the optical response in each direction is further emphasized by the polar plots in Fig.~\ref{Fig_Cd3As2_Na3Bi_sigma_vs_dir}(c)-(d), and (f)-(g), which show the dependence of the magnitude of the conductivity \cite{rashba}
		\begin{equation}
		\sigma = \sigma_{xx}\cos^{2}\varphi\sin^{2}\vartheta + \sigma_{yy}\sin^{2}\varphi\sin^{2}\vartheta + \sigma_{zz}\cos^{2}\vartheta
		\end{equation}
		on the polarization angles $\vartheta$ and $\varphi$ of some arbitrarily oriented driving field with respect to the major axes. Here, we define $\sigma$ such that it obeys $\mathbf{J} = \hat{\mathbf{r}}\sigma\lvert\mathbf{E}\rvert$, where $\mathbf{J}$ is the current density, $\mathbf{E}$ is the driving electric field, and the unit vector is $\hat{\mathbf{r}}\parallel\mathbf{J}$. 
	As the $\mathrm{Re}(\sigma_{ii})$ and $\mathrm{Im}(\sigma_{ii})$ share the same dependence on direction, we only plot the real part.  With reference to Eqs.~\ref{eqn_intraband_sigma} and \ref{eqn_interband_sigma}, we find that the strongly anisotropic response originates from the following scaling relation: $\sigma_{ii} \propto v_{i}/v_{j}v_{k}$ where $i\neq j \neq k$. This implies that if the Fermi velocities perpendicular to the direction of polarization are smaller than the Fermi velocity parallel to the direction of polarization, the optical response is enhanced. For DSMs such as those mentioned above, it is typical that $v_{x} \sim v_{y}$, and $v_{z} \ll v_{x},v_{y}$, indicating that for polarization along $x$ or $y$, the optical response will be the stongest. This is in agreement with our results in Fig.~\ref{Fig_Cd3As2_Na3Bi_sigma_vs_dir}, which clearly show the strongest optical response along the direction of the largest Fermi velocity (along $x$ for $\mathrm{Na_{3}Bi}$ and along $y$ for $\mathrm{Cd_{3}As_{2}}$).  We characterize this strong directional dependence through an optical \emph{extinction ratio} between an $i$-directional and a $j$-directional linear response $\Lambda_{ij} = J_i(\omega) / J_j(\omega)$ where $J_{i(j)}(\omega) = \sigma_{ii(jj)}(\omega) E(t)$ is the magnitude of the $i$($j$)-directional optical current density when subjected to an external time-varying electric field $E(t)$, which exhibits the following  \emph{universal scaling relation}:
	\begin{equation}\label{eqn_optical_anisotropy_ratio}
	\Lambda_{ij} = \frac{v_{i}^{2}}{v_{j}^{2}}.
	\end{equation} 
	This ratio can be can be as large as $\Lambda_{yz}\approx 15.8$ for $\mathrm{Cd_{3}As_{2}}$ and $\Lambda_{xz}\approx 19.2$ for $\mathrm{Na_{3}Bi}$. The extinction ratios for other directions and the experimentally measured values of the Fermi velocities are presented in Table~\ref{tbl_anisotropy_ratio}. It should be noted that Eq. \ref{eqn_optical_anisotropy_ratio} is \emph{independent} of the optical frequency, temperature, Fermi level and defect scattering effects. Thus, the strong optical anisotropy of Cd$_3$As$_2$ and Na$_3$Bi is expected to persist broadly over the frequency windows as long as the optically excited electrons remained well-described by the 3D Dirac conic band structure. For Cd$_3$As$_2$, the energy scale of the Dirac cone reaches $\gg 200$ meV \cite{PhysRevMaterials.2.120302}. This suggests that the strong optical anisotropy of anisotropic 3D DSMs should cover over an ultrabroad frequency band from sub-THz (limited by low energy Fermi velocity renormalization induced by many-body effects \cite{MB, MB2}) to at least 50 THz.   

	\begin{table}[t]
		\caption{Strong optical anisotropy of 3D Dirac semimetals $\mathrm{Cd_{3}As_{2}}$ and $\mathrm{Na_{3}Bi}$. The definition of the optical extinction ratio $\Lambda_{ij}$ where $i\neq j$ as defined by Eq.~\ref{eqn_optical_anisotropy_ratio}. We considered experimentally measured values of the Fermi velocities~\cite{Liu2014a,Science_Na3Bi_discovery}.}
		\label{tbl_anisotropy_ratio}
		\begin{ruledtabular}
			\begin{tabular}{c|cccccc}
				Material & $v_{x}$ (m/s) & $v_{y}$ (m/s) & $v_{z}$ (m/s) & $\Lambda_{xy}$ & $\Lambda_{yz}$ & $\Lambda_{xz}$\\
				\hline
				$\mathrm{Cd_{3}As_{2}}$ & $1.28\times 10^{6}$ & $1.30\times 10^{6}$  & $3.27\times 10^{5}$ & 0.97 & 15.8 & 15.3\\
				$\mathrm{Na_{3}Bi}$ & $4.17\times10^{5}$ & $3.63\times10^{5}$ & $0.95\times10^{5}$ & 1.32 & 14.6 & 19.2 \\
			\end{tabular}
		\end{ruledtabular}
	\end{table}

	We now turn our attention to the dielectric function of anisotropic DSMs. We computed the diagonal (i.e., longitudinal) elements of the dielectric tensor from the conductivity as 
	\begin{equation}
	\epsilon_{ii}(\omega) = \epsilon_{\mathrm{bg},i} + \frac{\mathrm{i}\sigma_{ii}^{\mathrm{tot}}(\omega)}{\omega\epsilon_{0}}
	\label{eqn_dielectric constant}
	\end{equation}
	where $\epsilon_{\mathrm{bg},i}$ is the background dielectric constant along the $i$ direction, which depends on the plasma frequency along $i\in\{x,y,z\}$ $\omega_{\mathrm{p},i}$, and the effective background dielectric constant accounting for the interband transitions obtained experimentally $\epsilon_{\infty}$ . However, as $\epsilon_{\infty}$ for 3D DSMs is typically measured as an isotropic value, we assume the $\omega_{\mathrm{p},x} = \omega_{\mathrm{p},y} = \omega_{\mathrm{p},z} = \omega_{\mathrm{p}}$, which we compute as:
	\begin{equation}
	\frac{\hbar\omega_{\mathrm{p}}}{\mathcal{E}_{\mathrm{F}}} = \sqrt{\frac{n_{e}e^{2}}{m^{*}\epsilon_{\infty}\epsilon_{0}}} = \sqrt{\frac{2r_{s}g}{3\pi\epsilon_{\infty}}}
	\end{equation}
	where $n_{e} = gk_{\mathrm{F}}^{3}/(6\pi^{2})$ is the electron density per Weyl cone, $k_{\mathrm{F}}$ is the Fermi wavevector magnitude, $m^{*}$ is the effective mass, $r_{s} = e^{2}/(4\pi\epsilon_{0}\hbar \bar{v}_{\mathrm{F}})$ is the fine structure constant calculated using the geometric mean Fermi velocity $\bar{v}_{\mathrm{F}} = (v_{x}v_{y}v_{z})^{1/3}$. We then calculate $\epsilon_{\mathrm{bg},i}$ by finding the zeros of the real part of Eq.~\ref{eqn_dielectric constant}:
	\begin{equation}
	\epsilon_{\mathrm{bg},i} = -\mathrm{Re}\Bigg{[}\frac{i\sigma_{ii}^{\mathrm{tot}}(\omega_{\mathrm{p}})}{\omega_{\mathrm{p},i}\epsilon_{0}}\Bigg{]}
	\end{equation}
	where $\sigma_{ii}^{\mathrm{tot}}(\omega_{\mathrm{p}})$ is obtained numerically from Eqs.~\ref{eqn_intraband_sigma} and \ref{eqn_interband_sigma} using a cubic spline.  We consider $\epsilon_{\infty} = 13$~\cite{Huang2015}, which applies to both $\mathrm{Cd_{3}As_{2}}$ and $\mathrm{Na_{3}Bi}$. These values yield the following background conductivities: $(\epsilon_{\mathrm{bg},x},\epsilon_{\mathrm{bg},y},\epsilon_{\mathrm{bg},z})\approx (28.8, 29.7, 1.88)$ for $\mathrm{Cd_{3}As_{2}}$, and $(\epsilon_{\mathrm{bg},x},\epsilon_{\mathrm{bg},y},\epsilon_{\mathrm{bg},z})\approx (25.1, 19.0, 1.30)$ for $\mathrm{Na_{3}Bi}$. By setting $v_{x} = v_{y} = v_{z} = 10^{6}$ m/s, we recover the isotropic value $\epsilon_{\mathrm{bg}} \approx 12.0$ presented in~\cite{PhysRevB.93.235417}.

	\begin{figure}[t]
		\centering
		\includegraphics[width = 85mm]{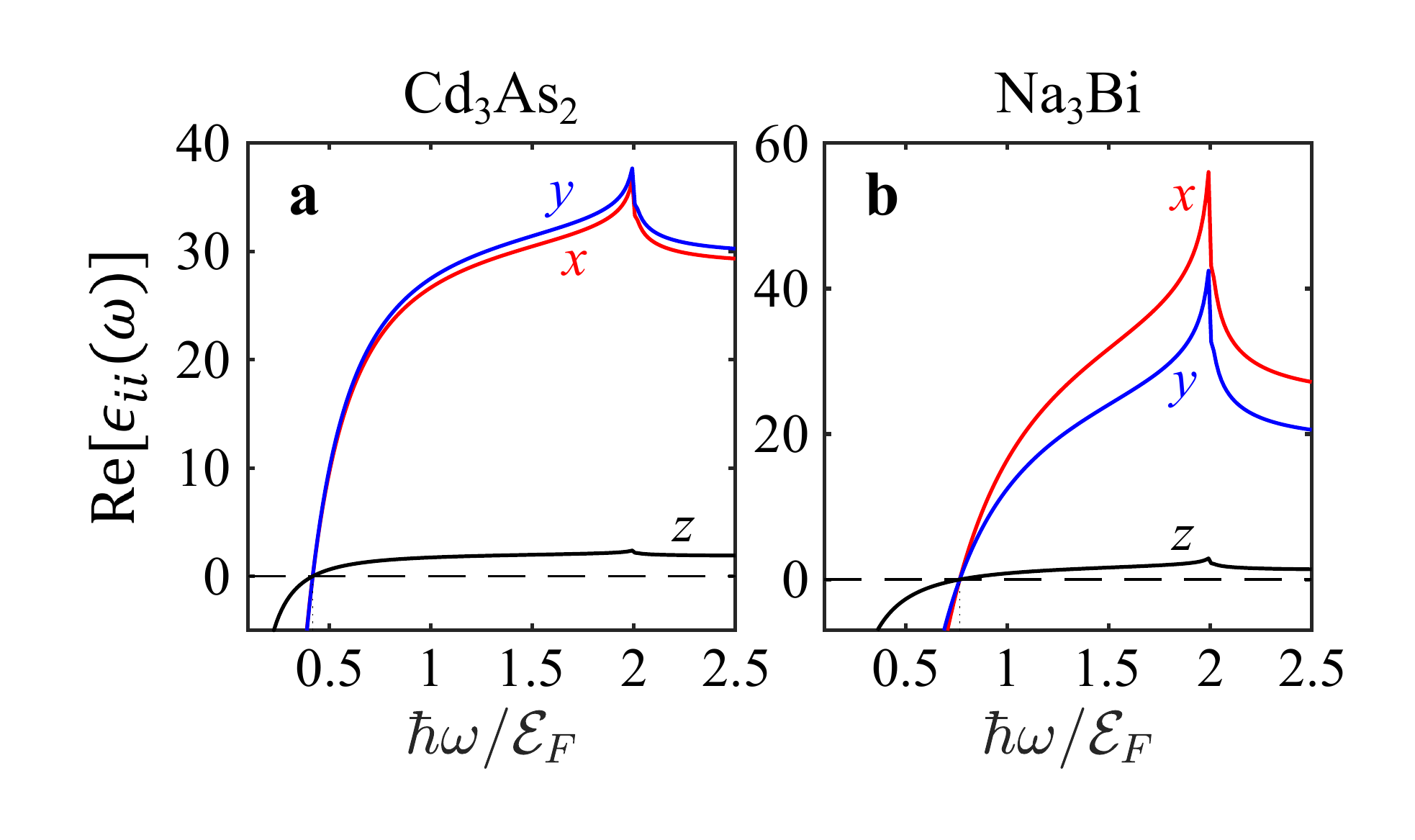}
		\caption{Real part of the dielectric function $\epsilon_{ii}(\omega)$ in each direction for 3D DSMs $\mathrm{Cd_{3}As_{2}}$ (a) and $\mathrm{Na_{3}Bi}$ (b). The plasma frequency in each direction $\omega_{\mathrm{p},i}$ correspond to $\mathrm{Re}[\epsilon_{ii}(\omega)] = 0$, indicated by the horizontal dashes in both panels. The plots above indicate the vast difference in the dielectric ($\mathrm{Re}[\epsilon_{ii}(\omega)] > 1$) or metallic ($\mathrm{Re}[\epsilon_{ii}(\omega)] < 0$) response of 3D DSMs between directions at a given frequency $\hbar\omega/\mathcal{E}_{\mathrm{F}}$. Unless otherwise stated, we consider the same parameters as  Fig.~\ref{Fig_Cd3As2_Na3Bi_sigma_vs_dir}}
		\label{Fig_re_epsilon_Na3Bi_Cd3As2}
	\end{figure}
	
	\begin{figure*}[t]
		\centering
		\includegraphics[width =158mm]{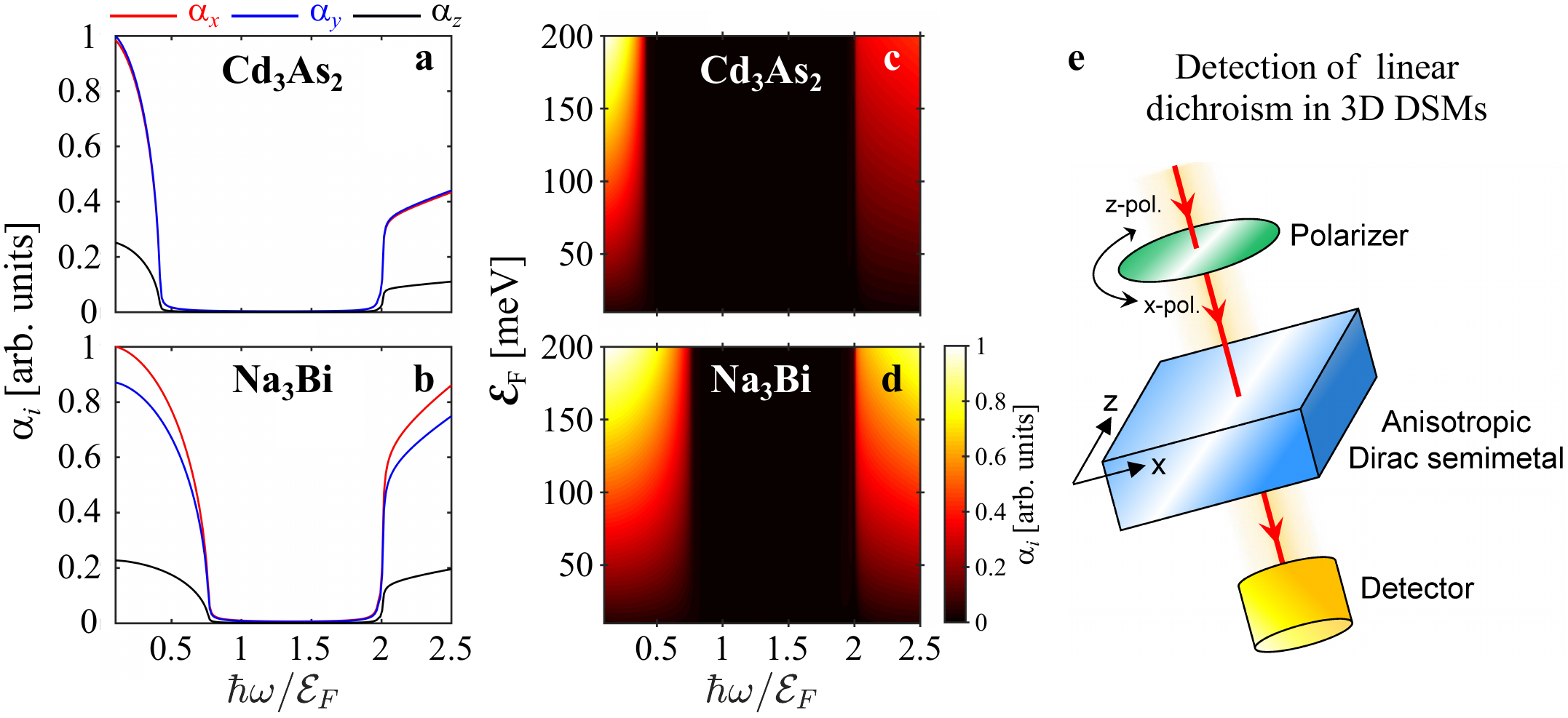}
		\caption{Linear dichroism of 3D DSMs $\mathrm{Cd_{3}As_{2}}$ and $\mathrm{Na_{3}Bi}$. We see from (a) and (b) that there is a large linear dichroism exhibited in the $z$-direction (black solid lines) due to the strong out-of-plane Fermi velocity anisotropy in $p_{z}$.  We show in (c) and (d) that the anisotropy of the aborption coefficient persists over a range of realistic Fermi levels. We find that the anisotropy ratio of the absorption coefficient, $\alpha_{i}/\alpha_{j} \propto v_{i}/v_{j}$. Unless otherwise stated, we consider the same parameters as in Figs.~\ref{Fig_Cd3As2_Na3Bi_sigma_vs_dir} and \ref{Fig_re_epsilon_Na3Bi_Cd3As2}. We compute the absorption coefficient using Eq.~\ref{eqn_abs_coeff}. We schematically illustrate in (e) a possible experimental geometry which could potentially enable the detection of the strong optical anisotropy of 3D DSM thin films. By changing the polarization of a linearly polarized laser normally incident on the $x$-$z$ plane of the 3D DSM thin film (since $v_{x}$ exceeds $v_{z}$ by a significant amount), the strength of the transmitted field detected will change.}
		\label{Fig_Abs_coeff_ratio_Na3Bi_Cd3As2_T4K_mu150meV_tau450fs}
	\end{figure*}
	
	We see from Fig.~\ref{Fig_re_epsilon_Na3Bi_Cd3As2} -- which shows the real part of $\epsilon_{ii}(\omega)$ for $\mathrm{Cd_{3}As_{2}}$ (a) and $\mathrm{Na_{3}Bi}$ (b), that when $\mathrm{Re}(\epsilon_{ii}) < 0$ ($\mathrm{Re}(\epsilon_{ii}) > 1$), 3D DSMs exhibit metallic (dielectric) response -- a behavior predicted~\cite{PhysRevB.93.235417} in isotropically dispersing 3D DSMs. However, we find once again that despite a qualitatively similar response in each direction, the magnitude in each direction possess the same $v_{i}/v_{j}v_{k}$ scaling as the conductivity. As a result, the anisotropy $\epsilon_{ii}/\epsilon_{jj}$ is also described by the universal scaling law given by Eq.~\ref{eqn_optical_anisotropy_ratio}. As 3D DSMs operating in the metallic regime can support the surface plasmon-polaritons (SPPs), the anisotropic dielectric function implies that the incident field polarization direction could potentially be used to tune the field confinement factor and propagation length of the SPPs. As opposed to isotropic DSMs such as graphene, this provides an additional degree of freedom for tuning the strength of light-matter interaction within 3D DSMs, which can potentially be a useful feature for plasmonic and sensing device applications.

	In the above analysis, we have considered an isotropic value of $\epsilon_{\infty}$ which is frequently measured in experiments. 
	We expect that should the measured value of $\epsilon_{\infty}$ be significantly different in each direction, the incident light polarization (with $\omega$ appropriately chosen) could tune not only parameters like the field confinement factor, but also serve as a switch between dielectric and metallic operation regimes. This arises as the zero-crossing $\mathrm{Re}(\epsilon_{ii}) = 0$ occurs at different values of $\omega_{\mathrm{p},i}$ for each direction $i$.  Hence anisotropy could additionally as serve a tuning parameter for the nature of light-matter interaction within 3D DSMs.

	We now calculate the absorption coefficient in the $i\in\{x,y,z\}$ Cartesian direction as~\cite{ZhaoNanoscale2018,LiuAdvOptMater2019}
	\begin{equation}
	\alpha_{i}(\omega) = \sqrt{2}\omega\Big{[}\sqrt{\epsilon_{ii}'(\omega)^{2} + \epsilon_{ii}''(\omega)^{2}} - \epsilon_{ii}'(\omega)\Big{]}^{1/2}.
	\label{eqn_abs_coeff}
	\end{equation}
	For compactness, we have defined $\epsilon_{ii}' = \mathrm{Re}(\epsilon_{ii})$ and $\epsilon_{ii}'' = \mathrm{Im}(\epsilon_{ii})$ , which we compute using Eq.~\ref{eqn_dielectric constant} and the numerical solution to the anisotropic linear conductivity $\sigma_{ii}(\omega)$.  Figure~\ref{Fig_Abs_coeff_ratio_Na3Bi_Cd3As2_T4K_mu150meV_tau450fs} shows the strong anisotropy of $\alpha_{i}(\omega)$ manifested as linear dichorism in $\mathrm{Cd_{3}As_{2}}$ (a,c) and $\mathrm{Na_{3}Bi}$ (b,d). In both cases, we find that in the frequency regimes $\omega \lesssim \omega_{\mathrm{p}}$ (metallic region) and $\omega > 2\mathcal{E}_{\mathrm{F}}/\hbar$ (beyond the Pauli-blocked region), the anisotropy in the absorption coefficient in different directions is the clearest. In the intermediate frequency regime  $\omega_{\mathrm{p}} < \omega < 2\mathcal{E}_{\mathrm{F}}/\hbar$, we see that $\alpha_{i}(\omega)$ is strongly quenched since $\epsilon_{ii}''(\omega) \approx 0$.  An inspection of Eq.~\ref{eqn_abs_coeff} reveals that the anisotropy ratio of the absorption coefficient between different directions scale as $\alpha_{i}/\alpha_{j} = v_{i}/v_{j}$ for all frequencies -- less drastic than the quadratic scaling of $\sigma_{ii}/\sigma_{jj}$ and $\epsilon_{ii}/\epsilon_{jj}$. However, despite the less pronounced anisotropy of the absorption coefficient, the linear dichroism of 3D DSMs can still be experimentally detected using the setup which we schematically illustrate in Fig.~\ref{Fig_Abs_coeff_ratio_Na3Bi_Cd3As2_T4K_mu150meV_tau450fs}(e). When the polarization of the laser normally incident on the plane of the 3D DSM thin film with the greatest Fermi velocity anisotropy is changed, the strength of the transitted fields should vary by a few times for both $\mathrm{Na_{3}Bi}$ and $\mathrm{Cd_{3}As_{2}}$ -- an easily detectable difference.

	%==================================================
	% Conclusion
	%==================================================
	\section{Discussion \& conclusion}

	We have shown, using the Kubo-Greenwood formula, that when anisotropic Fermi velocities in each direction are included, the optical response along each direction varies significantly. While the characteristic optical signatures of 3D Dirac electrons like the $\sigma(\omega) \propto \omega$ scaling is retained, our results show that the magnitude along each direction scales as $v_{i}/(v_{j}v_{k})$. This leads to the following universal scaling of the optical anisotropy ratio between the $i$ and $j$ directions: $\sigma_{ii}/\sigma_{jj} = \epsilon_{ii}/\epsilon_{jj} = (v_{i}/v_{j})^{2}$ -- a value which exceeds 15 times for $\mathrm{Cd_{3}As_{2}}$ and 19 times for $\mathrm{Na_{3}Bi}$, as shown in Table~\ref{tbl_anisotropy_ratio}. While we find that the qualitative trend of $\epsilon_{ii}(\omega)$ remains the same in all directions, the large anisotropy implies that for plasmonic applications, the polarization of the incident light can serve as an additional degree of freedom with which the field confinement factor and propagation length of SPPs can be tuned. In the case where $\epsilon_{\infty}$ are different in each direction, the plasma frequency $\omega_{\mathrm{p}}$ will acquire a directional dependence. This implies that at an appropriately chosen frequency, dielectric (e.g., waveguide modes) or metallic (SPPs) behavior can dominate depending on the direction of the incident light, thus opening up a novel device architecture where both the strength and the nature of light-matter interaction could be tailored using incident light polarization.
	We further remark that linear dichroism of 3D DSMs with Fermi velocities anisotropy is substantially stronger than many anisotropic optical materials as shown in Table~\ref{ext_compare} \cite{pbs,bp,sbse,ause,pero,1Dpero,gese,gese2,res2,tel,ant}.
	Although the extinction ratio is dwarfed by other exceptionally strong anisotropic optical materials, such as bilayer tellurene \cite{tel}, and antimonene, the broadband linear dichroism of Cd$_3$As$_2$ and Na$_3$Bi represents a unique strength not found in bilayer tellurene and antimonene.

	\begin{table}[t]
		\caption{Comparison of the extinction ratios of different anisotropic optical materials \cite{pbs,bp,sbse,ause,pero,1Dpero,gese,gese2,res2,tel,ant}.}
		\label{ext_compare}
		\begin{ruledtabular}
			\begin{tabular}{c c  c c }
				Material & Extinction ratio & Wavelength & Reference \\
				\hline
				$\mathrm{Cd_{3}As_{2}}$ & 15.8 & Broadband & This work \\
				$\mathrm{Na_{3}Bi}$		& 19.2 & Broadband & This work \\
				PbS nanowire 			& 2.38 &  532 nm	& \cite{pbs} \\
				Black phosphorus			& 5		& 1557 nm 	& \cite{bp} \\
				Sb$_2$Se$_3$			& 16	& 633 nm	& \cite{sbse} \\
				AuSe 					& 3 	& 450 nm 	& \cite{ause} \\
				(\textit{i}BA)$_2$(MA)Pb$_2$I$_7$ & 1.23 & 637 nm & \cite{pero} \\
				1D C$_4$N$_2$H$_14$PbI$_4$ & 5.5 & 405 nm 	& \cite{1Dpero} \\
				GeSe monolayer			& 3.4 	& 450 nm 	& \cite{gese}	\\
				GeSe nanoflakes			& 3.02	& 808 nm	& \cite{gese2} \\
				ReS2					& 3.5	& 532 nm 	& \cite{res2} \\
				Bilayer tellurene		& 2812	& 365 nm	& \cite{tel} \\
				Antimonene				& 145	& 387 nm	& \cite{ant}
			\end{tabular}
		\end{ruledtabular}
	\end{table}
	
	In summary, we studied the linear optical response of topological Dirac semimetal. 
	We found that the optical conductivity exhibits strong anisotropy with the universal scaling law, $\Lambda_{ij} = (v_i/v_j)^2$, independent of temperature, Fermi level and scattering effects, and is broadly applicable to the sub-THZ to at least 50 THz frequency window.  
	Recently, the optical properties of topological semimetals with nodal topology beyond Dirac semimetal, such as Weyl semimetal \cite{PhysRevB.101.085307} and nodal loop semimetal \cite{ChingHuaNodalHHG,PhysRevB.99.045124,NL, NL2}, have been extensively studied. We expect this ever-expanding family of topological semimetals \cite{ahn}, in which the optical and electronic properties are highly anisotropic along different crystal directions, to continually offer interesting platforms for the uncovering of exotic anisotropic optics and optoelectronic phenomenal, critical for the design of next-generation novel devices.  

	This work is supported by Singapore Ministry of Education (MOE) Tier 2 Grant (2018-T2-1-007) and USA ONRG grant (N62909-19-1-2047). JL is supported by MOE PhD RSS. KJAO acknowledges the funding support of Xiamen University Malaysia Research Fund, grant no. XMUMRF/2019-C3/IECE/0003 and XMUMRF/2020-C5/IENG/0025, and the Ministry of Higher Education Malaysia under the Fundamental Research Grant Scheme, grant no. FRGS/1/2019/TK08/XMU/02. CZ acknowledges the funding support by the Australian Research Council (Grant No. DP160101474).
	
	%	\bibliographystyle{apsrmp4-1} % uncomment to use bibtex
	%	\bibliography{anisotropic_3D_DSMs}  % uncomment to use bibtex

\end{document}